\documentclass[journal=jpclcd,manuscript=article]{achemso}

\usepackage[version=3]{mhchem} 
\usepackage{xcolor}
\usepackage{hyperref}
\usepackage{siunitx}
\DeclareSIUnit\photons{photons}
\DeclareSIUnit\Molar{M}
\DeclareSIUnit\wtpercent{wt\%}

\author{M. Moron}
\email{marc.moron@tu-dortmund.de}
\affiliation[TU Dortmund]
{Fakultät Physik und DELTA, TU Dortmund}

\author{A. Al-Masoodi}
\affiliation[Siegen University]
{Department Physik, Naturwissenschaftlich-Technische Fakultät, Universität Siegen }

\author{C. Lovato}
\affiliation[Siegen University]
{Department Physik, Naturwissenschaftlich-Technische Fakultät, Universität Siegen }

\author{M. Reiser}
\affiliation[Department of Physics, Stockholm University]
{Stockholm University, Stockholm, Sweden}

\author{L. Randolph}
\affiliation[Siegen University]
{Department Physik, Naturwissenschaftlich-Technische Fakultät, Universität Siegen }

\author{G. Surmeier}
\affiliation[TU Dortmund]
{Fakultät Physik und DELTA, TU Dortmund}

\author{J. Bolle}
\affiliation[TU Dortmund]
{Fakultät Physik und DELTA, TU Dortmund}

\author{F. Westermeier}
\affiliation[DESY]
{Deutsches Elektronen-Synchrotron DESY, Hamburg}

\author{M. Sprung}
\affiliation[DESY]
{Deutsches Elektronen-Synchrotron DESY, Hamburg}

\author{R. Winter}
\affiliation[TU Dortmund]
{Fakultät Chemie und Chemische Biologie, Physikalische Chemie, TU Dortmund}

\author{M. Paulus}
\affiliation[TU Dortmund]
{Fakultät Physik und DELTA, TU Dortmund}

\author{C. Gutt}
\email{gutt@physik.uni-siegen.de }
\affiliation[Siegen University]
{Department Physik, Naturwissenschaftlich-Technische Fakultät, Universität Siegen }

\title[An \textsf{achemso} demo]
  {Gelation dynamics upon pressure-induced liquid-liquid phase separation in a water-lysozyme solution}

\abbreviations{IR,NMR,UV}

\begin{document}

\begin{tocentry}




\includegraphics[width=5cm]{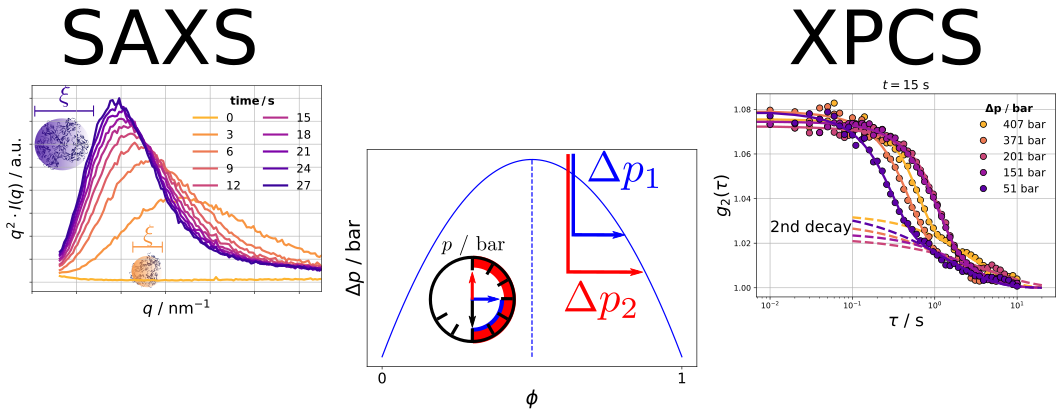}
\end{tocentry}

\begin{abstract}
Employing X-ray photon correlation spectroscopy we measure the kinetics and dynamics of a pressure-induced liquid-liquid phase separation (LLPS) in a water-lysozyme solution. Scattering invariants and kinetic information provide evidence that the system reaches the phase boundary upon pressure-induced LLPS with no sign of arrest. The coarsening slows down with increasing quench depths. The $g_2$-functions display a two-step decay with a gradually increasing non-ergodicity parameter typical for gelation. We observe fast superdiffusive ($\gamma \geq 3/2$) and slow subdiffusive ($\gamma < 0.6$) motion associated with fast viscoelastic fluctuations of the network and a slow viscous coarsening process, respectively. The dynamics age linear with time $\tau \propto t_\mathrm{w}$ and we observe the onset of viscoelastic relaxation for deeper quenches. Our results suggest that the protein solution gels upon reaching the phase boundary.
\end{abstract}

 Liquid-liquid phase separation (LLPS) provides a pathway for structure formation in biology via the formation of biomolecular condensates \cite{brangwynne2009germline,berry2018physical,shin2017liquid,hyman2011beyond}. The biological functions of these condensates—including steering biochemical reactions rates, sensing, or signaling—are being intensely investigated \cite{shin2017liquid}. 
  In the context of understanding biomolecular processes and life in the deep sea ocean or in the water-rich deep subsurface of planets\cite{fetahaj_biomolecular_2021}, pressure induced LLPS of protein solutions is highly relevant. Hydrostatic pressure allows to tailor the intermolecular interaction potential between proteins in a very mild way thus serving as an important physical probe for mapping their conformational landscape and phase behavior\cite{winter_interrogating_2019}. Furthermore, gelation and aggregation with subsequent fibril formation due to LLPS have been associated with a variety of diseases caused by a loss and/or change of function of the condensates \cite{malinovska2013protein}. 
 The state of the condensates depends on the dynamic and kinetic processes during their formation with dynamical asymmetries between the two phases producing a hierarchy of length and time scales and invoking viscoelastic properties of the resulting network structures \cite{berry2018physical,zaccarelli2007colloidal,tanaka2000viscoelastic}. 


Arrested phase separation has been observed for particles with short range attractive interactions such as colloids or small proteins \cite{Cardinaux2007,da2016kinetics,da2017arrested,Gibaud_2009,girelli2021}. However, the exact locations of glass and gel lines are still elusive as are the dynamic phenomena accompanied with the LLPS. For example, for deep temperature quenches of water-lysozyme \cite{Cardinaux2007} and BSA-water solutions \cite{da_vela_interplay_2020} the arrest line is located inside the spinodal region in agreement with predictions by mode-coupling theory for short-range attractive and long-range repulsive interaction potentials \cite{foffi2002phase}. In contrast, for colloidal particles with attractive interactions, the spinodal phase boundary has been found to coincide with the gel line identifying spinodal decomposition as driver for gelation  \cite{lu2008gelation}. 

Here, we employ X-ray photon correlation spectroscopy (XPCS) to investigate the dynamic processes during a pressure-induced LLPS of a lysozyme-water solution. The pressure-jump relaxation technique has several advantages over the temperature-jump relaxation technique for biomolecular systems: The final pressure is readily obtained so that sample inhomogeneity is no problem, the processes is bidirectional and fully reversible, and temperature-induced unfolding and irreversible aggregation of proteins is avoided. XPCS is the X-ray analogue of dynamic light scattering employing coherent X-rays capable of resolving the collective temporal heterogeneous dynamics on the required length scales, ranging simultaneously from nanometers to microns and timescales from microseconds to hours \cite{moller2019x,girelli2021,begam2021kinetics,perakis2020towards}. Reducing pressure (i.e. the repulsive part of the interaction) we quench the system into the metastable region of the phase diagram. From the scattering invariants we conclude that the system reaches the phase boundary accompanied by a slowing down of the coarsening process visible via a decrease of coarsening exponents from $1/3.2$ to values of $1/4.6$. 
The $g_2$ functions display a two-step decay indicating the initial surface formation (superdiffusive dynamics) and the onset of gelation (subdiffusive dynamics). Our results show that upon pressure release dense lysozyme solutions in water forms a protein network within a few seconds. This network forms a gel which consequently coarsens and ages over much longer time scales. The gel is a soft nanostructured network that can support small amounts of stress.

\begin{figure}[!htb]
    \centering
    \includegraphics[scale=0.85]{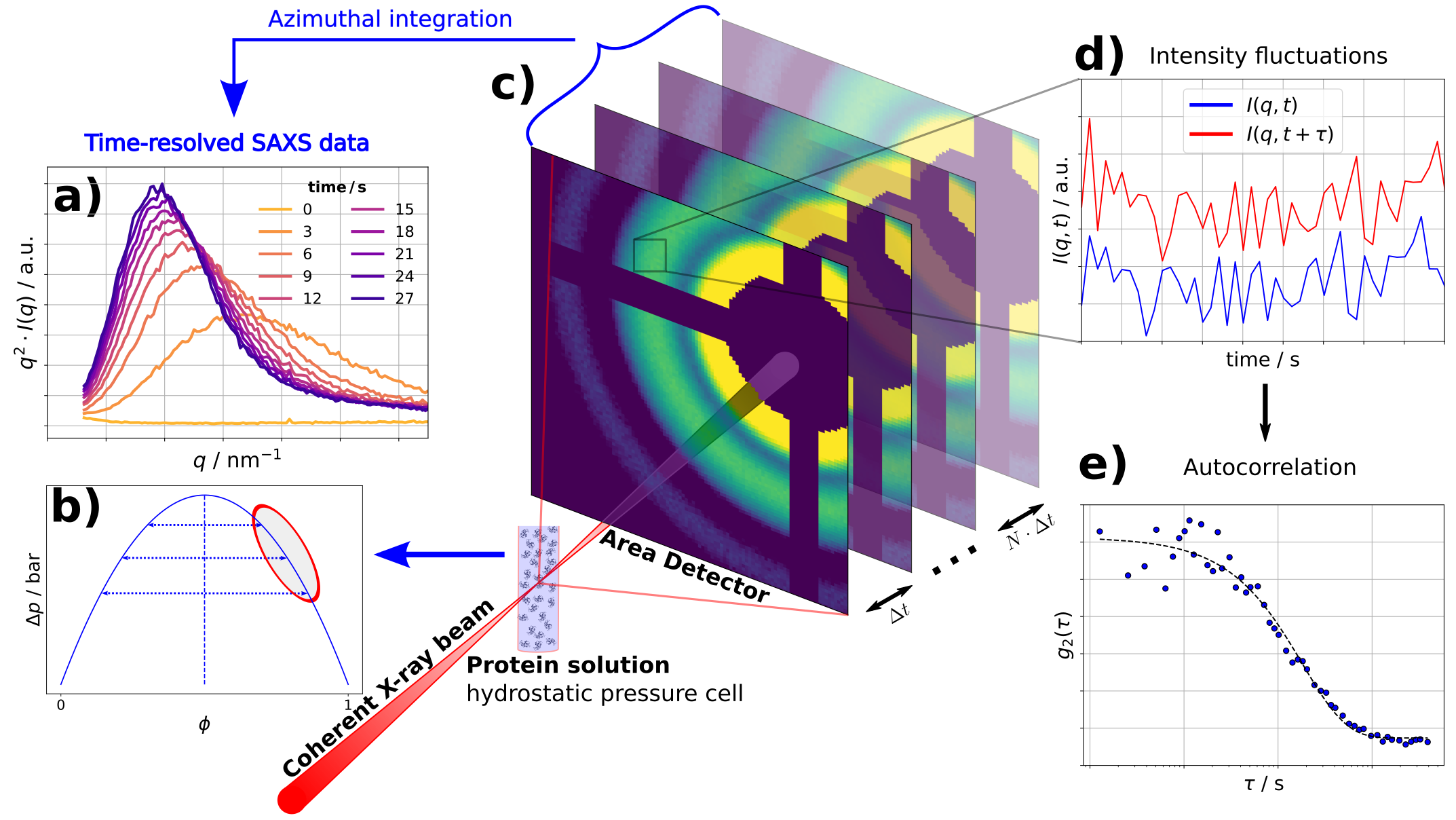}
    \caption{Schematics of the XPCS experiment at P10, DESY during a pressure-induced LLPS of a lysozyme-water solution (238 mg/mL, 500 mM NaCl). During the LLPS a series of X-ray speckle patterns is recorded (c). Azimuthal integration yields the time-resolved SAXS data displaying the growth of the correlation peak during LLPS (a). A correlation analysis yields the corresponding $g_2$ autocorrelation functions (d-e).}
    \label{fig:Explanation_XPCS}
\end{figure}

XPCS experiments were conducted at the P10 Coherence Beamline at PETRA III, Deutsches Elektronen-Synchrotron, employing an X-ray beam of photon energy $E=\SI{13}{\kilo\eV}$, a beam size of $100 \times 100$ $\si{\square\micro\meter}$, and a maximum  photon density of $\SI{e7}{photons\per\s\per\micro\square\meter}$. Time series of coherent diffraction patterns were collected with an EIGER 4-mega-pixel detector at a distance of \SI{21.2}{\meter}, covering a q-range from \SIrange{0.006}{0.5}{\per\nano\meter}. Hen egg white lysozyme was purchased from Sigma Aldrich (product code: 10837059001) and used without further purification. We dissolved the protein in $\SI{25}{\milli\Molar}$ BisTris buffer (pH7) to a concentration of $\SI{13}{\wtpercent}$. We then dialyzed the protein solution to obtain higher protein concentrations (see \cite{schulze_phase_2016} for details). The dialyzed lysozyme stock solution (\SI[per-mode=symbol]{286}{\milli\gram\per\milli\liter}) was solved in a \SI{25}{\milli\Molar} BisTris buffer solution (pH7) containing \SI{3}{\Molar} NaCl in a ratio of 1:5 leading to a final protein concentration of \SI[per-mode=symbol]{238}{\milli\gram\per\milli\liter} and NaCl concentration of \SI{500}{\milli\Molar}. A home-built hydrostatic pressure cell (details in \cite{krywka2008}) with diamond windows was used for pressurizing the sample and for control of the sample temperature ($T=\SI{280}{K}$). 
After loading, the sample is pressurized to $\SI{1}{\kilo\bar}$ into the homogeneous area of the phase diagram. Having reached equilibrium a series of X-ray exposures is started while simultaneously releasing the pressure. At p=$\SI{700}{\bar}$ an intense SAXS pattern indicates the onset of the LLPS (schematic setup Fig.\ref{fig:Explanation_XPCS}). In the following the quench depths $\Delta p$ are measured with respect to $p=\SI{700}{\bar}$.
An azimuthal integration yields the time depending SAXS patterns, while two-time intensity correlation functions (TTCs) for specific wave vectors $q$ and different quench depths $\Delta p$ are calculated via \begin{equation}
    C(t_1, t_2, q) = \frac{\langle I(q,t_1)I(q,t_2) \rangle}{\langle I(q,t_1) \rangle\langle I(q,t_2)\rangle},
\end{equation}
with $\langle ... \rangle$ being an average over pixels corresponding to a specific q-range.
Waiting time $t_\mathrm{w}$ dependent $g_2(\tau,t_\mathrm{w},q)$ functions have been extracted by diagonal cuts with $\tau=t_1-t_2$ along the TTC as indicated in Fig. \ref{fig:Dynamics}(a).




At high pressures of $\SI{1}{\kilo\bar}$ and low temperatures ($T=\SI{280}{\K}$) the electrostatic repulsive part of the protein-protein interactions and full hydration of the protein interface (avoiding formation of transient void volume at contact interfaces) stabilize the homogeneous lysozyme solution. Upon pressure reduction the short range attraction induces the LLPS visible by the intense time depending growth of the SAXS signal (Fig.\ref{fig:Explanation_XPCS}(a)). We observe the typical X-ray scattering fingerprints of the LLPS, a growth of the SAXS peak in intensity and a shift to smaller q-values as time progresses on time scales of a few seconds. The maximum of the scattering intensity increases with quench depth $\Delta p$ due to higher concentration and volume fraction of the dense lysozyme phase. 

\begin{figure}[!htb]
    \centering
    \includegraphics[scale=.8]{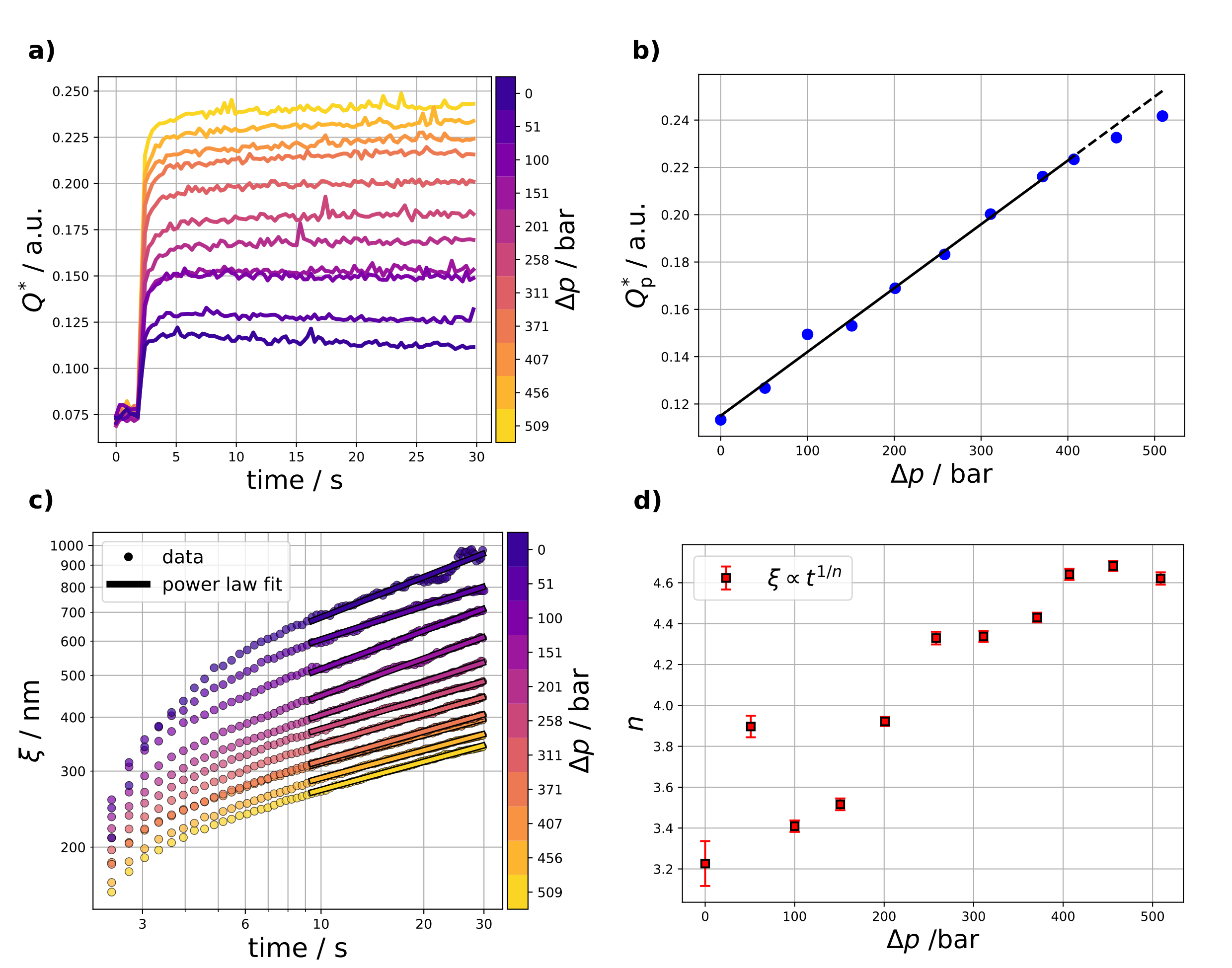}
    \caption{a) Time evolution of the scattering invariant $Q^*$ for all measured quench depths $\Delta p$. b) Plateau value of the scattering invariant $Q^*_{\mathrm{p}}$ versus quench depth. c) Time evolution of the correlation length $\xi=2\pi / q_{\mathrm{max}}$. Solid lines represent power law fits $\xi \propto t^{(1/n)}$, d) resulting power law exponent $n$.}
    \label{fig:SAXS-ana}
\end{figure}

\begin{figure}[!htb]
    \centering
    \includegraphics[scale=0.8]{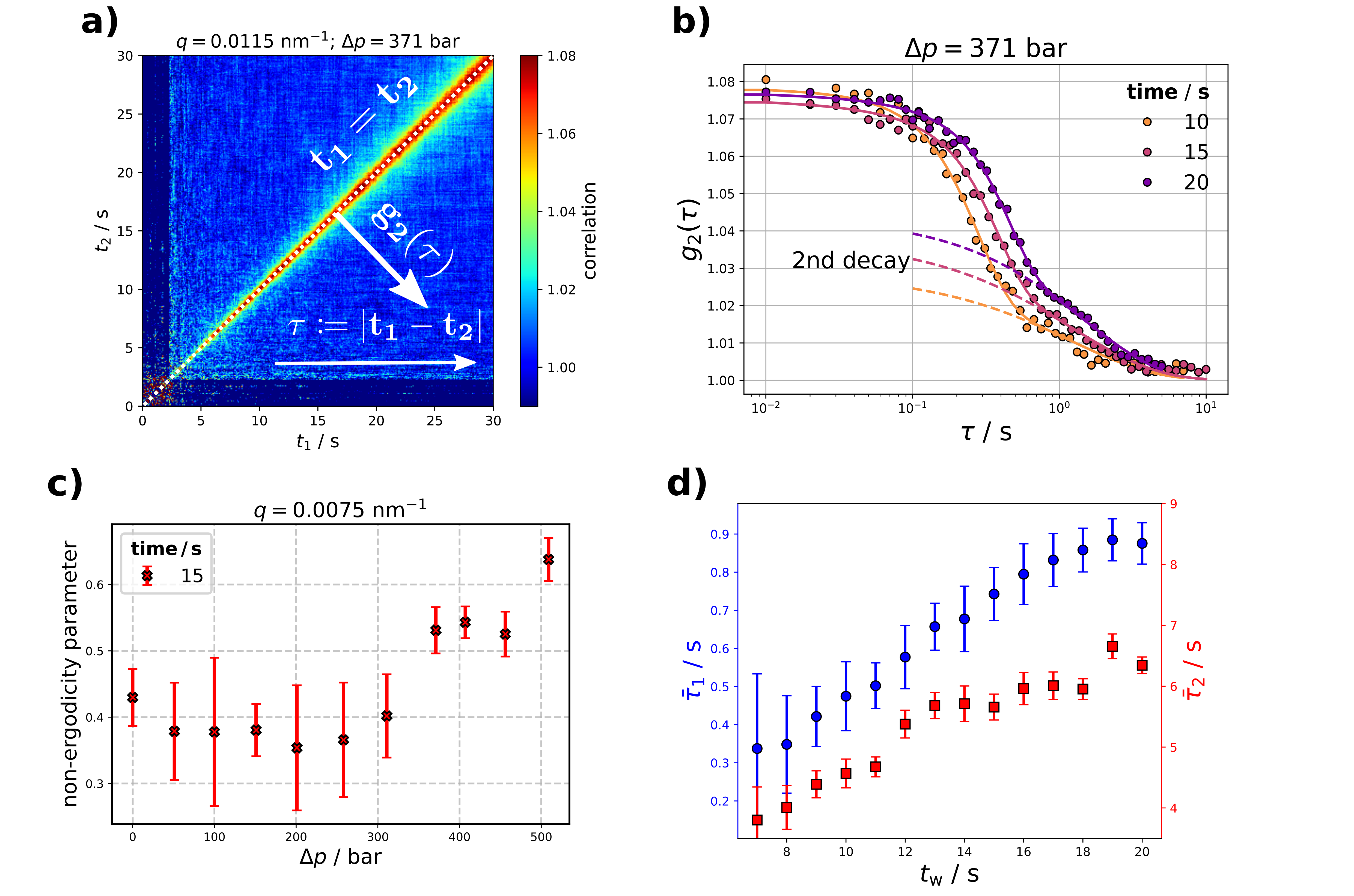}
    \caption{Correlation analysis of LLPS of a lysozyme-water solution for $\Delta p = \SI{371}{\bar}$. a) Two-time correlation map during quench at $q=\SI{0.0115}{\per\nano\meter}$. b) $g_2$ functions for different waiting times $t_{\mathrm{w}}$. The solid lines represent the KWW fits and the dashed lines indicate the contribution of the slow decay. c) Non-ergodicity parameter $f$ in dependence on quench depth for $t_{\mathrm{w}}=\SI{15}{\s}$ at $q = \SI{0.0075}{\per\nano\meter}$. d) waiting time dependence of the relaxation time for the fast ($\tau_1$, blue) and slow ($\tau_2$, red) process (averaged for $\Delta p$ in the range of \SIrange{407}{509}{\bar}).}
    \label{fig:Dynamics}
\end{figure}

We first evaluate the scattering invariant $Q^{*}= \int dq q^2 I(q)$ for different quench depths (see Fig. \ref{fig:SAXS-ana}(a)). Recently, Da Vela \cite{da_vela_interplay_2020} suggested to use the quench depth dependence of the invariant as a sensitive marker for the system reaching the binodal phase boundary upon LLPS. Following this methodology and plotting the plateau value of the invariant $Q^*_{\mathrm{p}}$ (reached after \SI{7}{\s}) as a function of quench depth $\Delta p$ we observe a linear increase of $Q^*_{\mathrm{p}}$ (see Fig. \ref{fig:SAXS-ana}(b)) indicating that for all of our pressure quenches the system reaches the binodal phase boundary and does not arrest before upon crossing a glass line. 

We verify this independently by fitting the early growth of the scattering intensity to $\ln{I(q,t)} \propto 2R(q,\Delta p) \cdot t$ with a growth rate $R(q,\Delta p)$ that depends on $q$ and quench depth. Making use of expressions from the linearized Cahn-Hilliard equation we fit the rates to $R(q) =D_{0}q^2 (1-(q/q_{\mathrm{c}})^2)$ with the diffusion constant $D_0$ and the critical wavevector $q_{\mathrm{c}}$ as refinement parameters. The diffusion constant decreases continuously with quench depth from values of $D_0= \SI[separate-uncertainty=true]{7600 +- 150}{\square\nano\meter\per\s}$ ($\Delta p=\SI{50}{\bar}$) to $D_0=\SI[separate-uncertainty=true]{2600 +- 170}{\square\nano\meter\per\s}$ ($\Delta p=\SI{509}{\bar}$) with no indication of arrest. We determine the mean domain size $L$ for each quench in the time interval 2.4 to 5 s and estimate the viscosity via $\eta(\Delta p) = k_{\mathrm{B}}T/(6 \pi L D_0)$ and the concentration $\phi$ via $\eta=\eta_{H2O}(1-\phi/\phi_{\mathrm{G}})^{2.1}$ with $\phi_{\mathrm{G}}$ being the concentration at the glass transition. We obtain concentrations of $\phi =0.85 \cdot \phi_{\mathrm{G}}$ for quench depth $\Delta p=\SI{50}{\bar}$ up to $\phi =0.94 \cdot \phi_{\mathrm{G}}$ at $\Delta p=\SI{509}{\bar}$. While the absolute numbers of the concentrations need to be taken with some caution
the trend confirms the finding that the system does reach the boundary of the phase diagram.

Previous work on lysozyme solutions reported a crossing of the glass line upon LLPS for low temperature quenches below an arrest tie line \cite{Cardinaux2007,Gibaud_2009}. We note that in terms of location in the phase diagram the quench depth here is quite shallow as a pressure jump of \SI{1}{\kilo\bar} corresponds to only 1 $k_{\mathrm{B}}T$ change in interaction energy \cite{schroer_nonlinear_2011}. Thus we are exploring the topmost part of the spinodal region close to the critical point. 

The SAXS intensity shows that for shallow quenches the maximum $q_{\mathrm{max}}$ moves continuously towards smaller values whereas for deeper quenches after an initial growth the peak evolution slows down. The growth kinetics of the characteristic length $ \xi =2 \pi/q_{\mathrm{max}}$ is shown for selected quench depths in Fig. \ref{fig:SAXS-ana}(c).
Typical length scales range between \SIrange{180}{300}{\nano\meter} three seconds after quench increasing to values of 280 nm ($\Delta p = \SI{500}{\bar}$) up to \SI{700}{\nano\meter} ($\Delta p = \SI{0}{\bar}$) ten seconds after quench. For very long waiting times on the order of hundreds of seconds the maximum of the SAXS intensity moves out of the observable q-window indicating length scales of \SI{1}{\micro\meter} and above. It is clearly seen that for all quenches the growth of the characteristic length $\xi$ follows a power law $\xi \sim t^{1/n} $. For shallow quenches we find $n \approx 3.2$ which is consistent with coarsening mechanisms based on diffusion or coalescence \cite{siggia1979late}. With increasing quench depth, the growth of $\xi$ slows down as indicated by the increase of $n$ to values of 4.6. For shallow quenches the surface tension is the main driving force for the domain coarsening while for deeper quenches surface tension becomes more and more unable to advance the coarsening process, because the dense phase has a high viscosity and becomes viscoelastic \cite{testard2014intermittent}. As a consequence, surface tension is no longer able to relax in a significant manner the curved interfaces formed during the phase separation process leading to an increase of $n$. This has also been predicted by simulations when coupling the LLPS with a reduced mobility and gelation in the dense phase due to concentration \cite{sciortino1993interference,Testard}. 

We note that beyond a quench depth of \SI{250}{\bar} $n$ only slightly increases from values of 4.25 to 4.6 for quench depths up to 500 bar (Fig.\ref{fig:SAXS-ana}(d)). This behavior is in contrast with deep temperature quenches for BSA or IgG systems in which considerable larger $n$-values have been reported \cite{da2016kinetics,da2017arrested} indicating an almost complete arrest of coarsening.

The dynamics of the system during the LLPS can be followed by the intensity correlation functions. Around \SI{2.5}{\s} after quenching the TTCs start to display a finite correlation value above the noise (Fig.\ref{fig:Dynamics}(a)).
The width of the red line visible in the TTCs serves as an indicator of the typical main decorrelation time of the system. Thus, we see that the system slowly ages (Fig.\ref{fig:Dynamics}(a)) (i.e. slower correlation times) after the pressure quench. However, in contrast to a recent experiment on temperature quenches on IgG/PEG \cite{girelli2021}, the TTCs for lysozyme-water do not display an onset of a much broader frozen in area within the first \SI{900}{\s} (SI Fig.\ref{fig:TTC-long}) and a fast decorrelation channel is always visible.

The $g_2$ functions obtained from diagonal cuts are shown in Fig.\ref{fig:Dynamics}(b) for a quench depth of $\Delta p=\SI{371}{\bar}$. We observe a two step decay of the $g_2$ functions with the second decay being more pronounced for deeper quenches. We attribute the fast decay to the interface formation and its successive fluctuations and the slower decay to the coarsening process. Both processes slow down with waiting time with the slow process only slightly increasing in relative contribution to the overall decorrelation. We observe dynamical heterogeneities and fluctuations in the decorrelation times (SI Fig.\ref{fig:ttc_and_tau}) as they often occur in out-of-equilibrium processes \cite{Duri_2006}. By averaging decorrelation times for deeper quenches we can clearly identify the aging of the system for both processes (Fig. \ref{fig:Dynamics}(d)). Aging and the presence of a non-ergodicity parameter demonstrate the onset of the gelation process \cite{cipelletti2003universal,zaccarelli2007colloidal}.   

Fig.\ref{fig:nonergo}(b) displays the $g_2$ functions at a waiting time $t_{\mathrm{w}}=\SI{15}{\s}$ for different quench depths. A non-monotonic behavior of decorrelation times with $\Delta p$ is apparent. The shallow quench to $\Delta p=\SI{51}{\bar}$ displays the fastest dynamic while quenches to \SI{151}{\bar} and \SI{201}{\bar} are much slower. Even deeper quenches are again faster (\SI{371}{\bar} and \SI{407}{\bar}). The $g_2(t,t_{\mathrm{w}},q)$ functions have been fitted by a sum of Kohlrausch-Williams-Watts (KWW) functions 
\begin{equation} \label{eq:kww-model}
g_2(t,t_{\mathrm{w}},q)=A_1\exp[-(2\Gamma_1t)^{\gamma_1}] +A_2\exp[-(2\Gamma_2t)^{\gamma_2}]
\end{equation}
with q-dependent relative amplitudes $A_1$,$A_2$, rates $\Gamma_1$ and $\Gamma_2$ and KWW exponents $\gamma_1$ and $\gamma_2$. Both rates differ by at least an order of magnitude and both decrease 
linearly with waiting time $t_{\mathrm{w}}$. At very long delay times of \SI{300}{\s} and beyond the slow processes fall out of equilibrium reaching time scales of up to \SI{1}{\hour}, while the fast process still decorrelates and ages at late time with $\tau_1 \sim  0.15 \cdot t_{\mathrm{w}}$ at $\Delta p=\SI{325}{\bar}$.

The non-ergodicity parameter $f(q)=A_2/(A_1+A_2)$ increases notably with quench depth (Fig. 3(c)) reflecting the successively pronounced gelation for higher concentrations. From the q-dependence of $f(q)$ at $\Delta p=\SI{509}{\bar}$ (Fig. S2(a)) we estimate a value of the localization length of $\delta \approx \SI[separate-uncertainty=true]{225 +- 10}{\nano\meter}$ using the Debye-Waller factor expression $f_q=\exp(-q^2\delta^2/6)$. 
Using  $G_0 \approx 2 k_{\mathrm{B}}T/(\delta^2 L)$ \cite{krall1998internal}, with $L=\SI{500}{\nano\meter}$ as typical length scale, we obtain an estimate for the value of the plateau modulus $G_0\approx\SI{0.4}{\pascal}$. The values at smaller quench depths are lower by factor of two and we note that these values of $G_0$ are quite typical for soft physically linked gels such as hydrogels for example \cite{jaspers2014ultra}.

\begin{figure}[!htb]
    \centering
    \includegraphics[scale=.8]{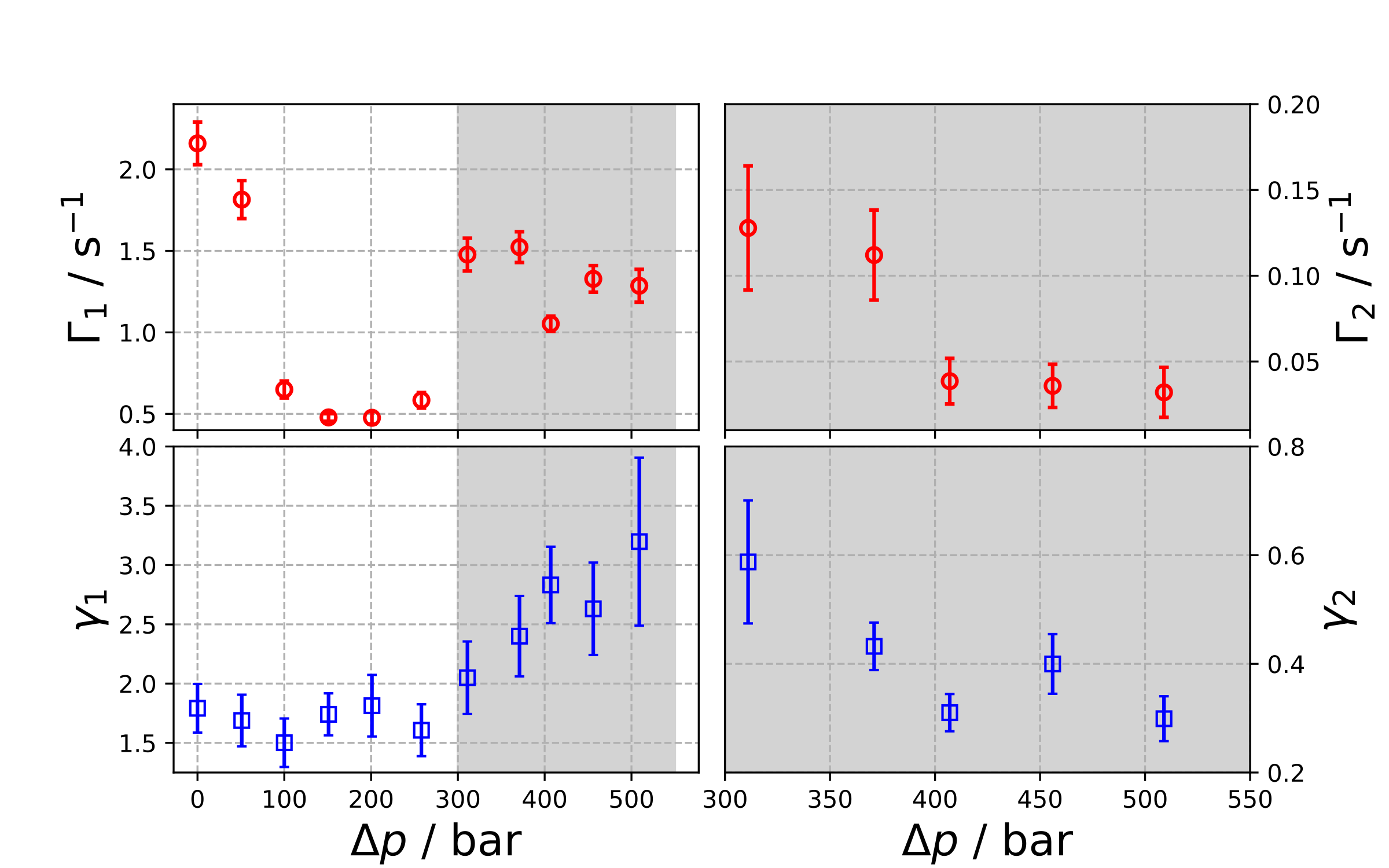}
    \caption{Left: Relaxation rate $\Gamma_1$ and KWW exponent $\gamma_1$ of the fast process. Right $\Gamma_2$, $\gamma_2$ for the slow process for $t_{\mathrm{w}}=\SI{15}{\s}$ and $q = \SI{8e-3}{\per\nano\meter}$.}
    \label{fig:gamma-kww}
\end{figure}

Fig.\ref{fig:gamma-kww} displays the pressure dependence of relaxation rates $\Gamma$ and the corresponding KWW exponents $\gamma$. We observe a  rather non-monotonic behavior especially for the fast rate $\Gamma_1$. The dynamics $\Gamma_1$ is fastest for low pressure quenches and then slows down reaching a minimum at $\Delta p=\SI{200}{\bar}$ and then speeding up again before reaching a plateau at the largest quench depths. 
The q-dependence of the rate of the fast process displays a ballistic type of motion $\Gamma_i \propto q$ (see SI Fig.\ref{fig:gamma_vs_q}). In contrast, we observe an almost flat dispersion for the slow decay which we attribute to the influence of the structure factor $S(Q)$ and the hydrodynamic interactions $H(Q)$ both altering the collective diffusion $D(Q) \propto H(Q)/S(Q)$ \cite{Banchio2008}.


The values of the KWW exponent of the fast mode indicate a change of dynamic properties for deeper quenches with $\gamma_1 \approx 3/2$ for low quenches and then increasing to values of \SIrange{2.5}{3}{} for quench depths of $\Delta p= \SI{300}{\bar}$ and beyond. Compressed exponents with values between \SIrange{1.5}{2.5}{} are in good agreement with recent numerical simulations of correlation functions during an arrested LLPS based on the Cahn-Hilliard equation \cite{girelli2021}. We attribute the increase of $\gamma_1$ at deeper quenches to the onset of viscoelastic behavior as regularly observed in gel-forming soft matter systems \cite{cipelletti2003universal,Duri_2006}, in agreement with the parallel increase of the non-ergodicity parameter at the same quench depths(Fig. 3(c)). Assuming a Maxwell model and making use of the deduced value of viscosity we estimate the viscoelastic relaxation time to be on the order of $\tau_0 = \eta / G_0 \approx\SI{1}{\s}$ ($\Delta p=\SI{509}{\bar}$) supporting the notion of increasingly viscoelastic behavior of the fast decay with deeper quenches.   

The peculiar ballistic type of motion $\Gamma \propto q$ and KWW exponents of $\gamma=3/2$ have been phenomenologically   explained in a mean field model with the motion of droplets under the action of internal stress \cite{cipelletti2003universal} in which two protein droplets tend to fuse and restructure locally the environment. The result of their motion is a force dipole with an elastic strainfield $u(r,t) \propto t \cdot  1/r^{\alpha}$ ($\alpha=2$ for dipole) which evolves linear in time until the dipole collapses. The resulting probability distribution function ($pdf$) of the strain $u$ scales as $pdf \propto u^{-3/\alpha-1=-5/2}$ and the KWW exponent as $\gamma=3/\alpha = 1.5$. In this mean field scenario, the observed increase of the KWW exponent implies values of $\alpha<2$ indicating a longer range of the strain field than predicted by the model of force dipoles. The origin of this behavior is presently unknown. Possible explanations comprise an increase in local rigidity as reported in simulations of clusters in metallic glasses with KWW exponents between 1.5 and 3 \cite{wu2018stretched}, or the onset of spatial-temporal correlations in models of continuous time random walks (CTRW) \cite{tejedor2010anomalous}. Nevertheless, it seems that this gradual change in dynamics is connected  with the lysozyme network slowing down and the dynamics becoming progressively more viscoelastic as the Maxwell relaxation time $\tau=\eta / G_0$ increases.


In contrast to the fast motion, the second and much slower decay shows a pronounced subdiffusive behavior with $\gamma_2$ decreasing from values of 0.6 ($\Delta p=\SI{309}{\bar}$) to 0.3 ($\Delta p=\SI{509}{\bar}$). As this decay is slower than the typical Maxwell relaxation time it is viscous in nature. We attribute this increase in degree of subdiffusivity to increasing constraints in the displacements of the network resulting in a slowdown of its dynamics and induced greater heterogeneity in relaxation times \cite{Cho}. The stretched exponential relaxation of colloidal gels is generally understood as a superposition of multiple overdamped modes, each exhibiting a single exponential decay \cite{Krall}. In this framework, the deeper quenches slow down the system and the distribution of characteristic timescales of the exponentials broadens due to dynamical heterogeneities. Recently, a simple connection of $n=\gamma/3$ between the coarsening exponent $n$ and the KWW coefficient $\gamma$ of droplet motion has been put forward \cite{lee_chromatin_2021}. In this scenario the slow down of coarsening of droplets in an embedded condensate is connected with the onset of subdiffusive ($\gamma<1$) behavior of the droplets. Based on this argument we would expect values of $n$ between 5 and 10 for the lowest quench which need to be compared to measured values of 4 - 4.5 for the deepest quenches, leaving us with a qualitative agreement only.

In conclusion, we measured the dynamics during pressure induced LLPS in a water lysozyme solution. The scattering invariants and the diffusion constants suggest that the system reaches the binodal phase boundary upon LLPS. The XPCS results demonstrate that a gelation process sets in upon phase separation with a more pronounced slow down and increasing non-ergodicity parameter for deeper quenches. 
A fast superdiffusive viscoelastic relaxation attributed to localized network fluctuations is observed together with a slow subdiffusive viscous relaxation from the coarsening dynamics.  
Gelation due to phase separation at the phase boundary has been predicted as a universal property of particles with short range attractions \cite{lu2008gelation}. Our results are in agreement with the notion that for shallow quenches the phase boundary of a water-lysozyme solution coincides with the gelation line. Thus simple globular proteins such as lysozyme are capable of forming soft nanostructured protein gels upon reducing the repulsive part of the interaction potential. 


\bibliography{protein_pressure}

\begin{acknowledgement}
C.G. acknowledges BMBF (Grants No. 05K19PS1 and No. 05K20PSA) for financial support.
Gefördert durch die Deutsche Forschungsgemeinschaft (DFG) im Rahmen der Exzellenzstrategie des Bundes und der Länder – EXC 2033 – Projektnummer 390677874.
Funded by the Deutsche Forschungsgemeinschaft (DFG, German Research Foundation) under Germany´s Excellence Strategy – EXC 2033 – Projektnummer 390677874.
\end{acknowledgement}
\newpage
\begin{suppinfo}
\setcounter{figure}{0}
\makeatletter
\renewcommand{\thefigure}{S\@arabic\c@figure}
\makeatother






\begin{figure}[!htb]
    \centering
    \includegraphics[scale=.7]{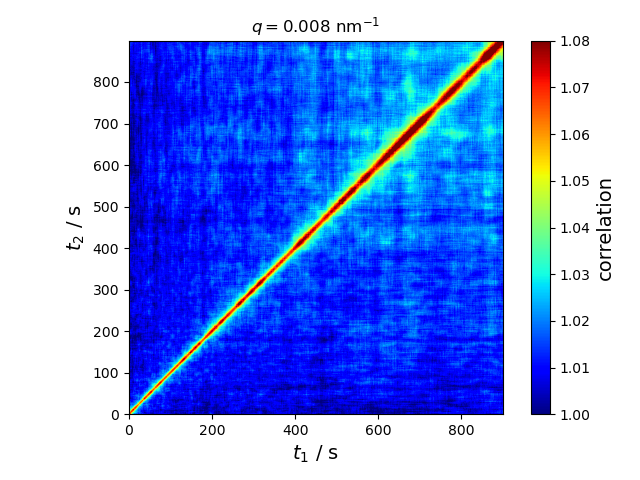}
    \caption{TTC starting \SI{30}{\s} after quenching ($\Delta p = \SI{325}{\bar}$) at $q=\SI{0.008}{\per\nano\meter}$ for a \SI{238}{\milli\gram\per\milli\liter} lysozyme sample.}
    \label{fig:TTC-long}
\end{figure}

\begin{figure}[!htb]
    \centering
    \includegraphics[scale=.8]{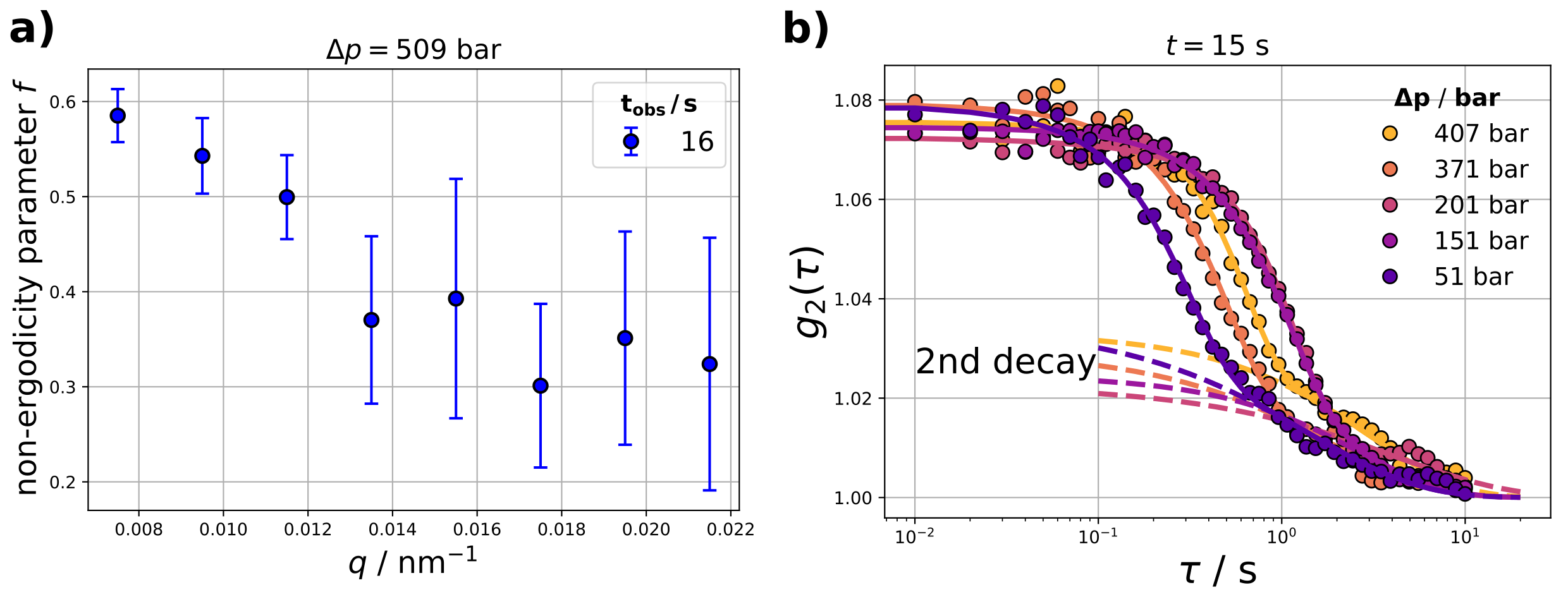}
    \caption{a) $q$-dependence of the non-ergodicity parameter $f_q$ extracted from the KWW fit for $\Delta p=\SI{509}{\bar}$ at $t_{\mathrm{w}}=\SI{16}{\s}$. b) $g_2$ functions for different quench depths $\Delta p$ at $q=\SI{0.0095}{\per\nano\meter}$.}
    \label{fig:nonergo}
\end{figure}

\begin{figure}[!htb]
    \centering
    \includegraphics[scale=.9]{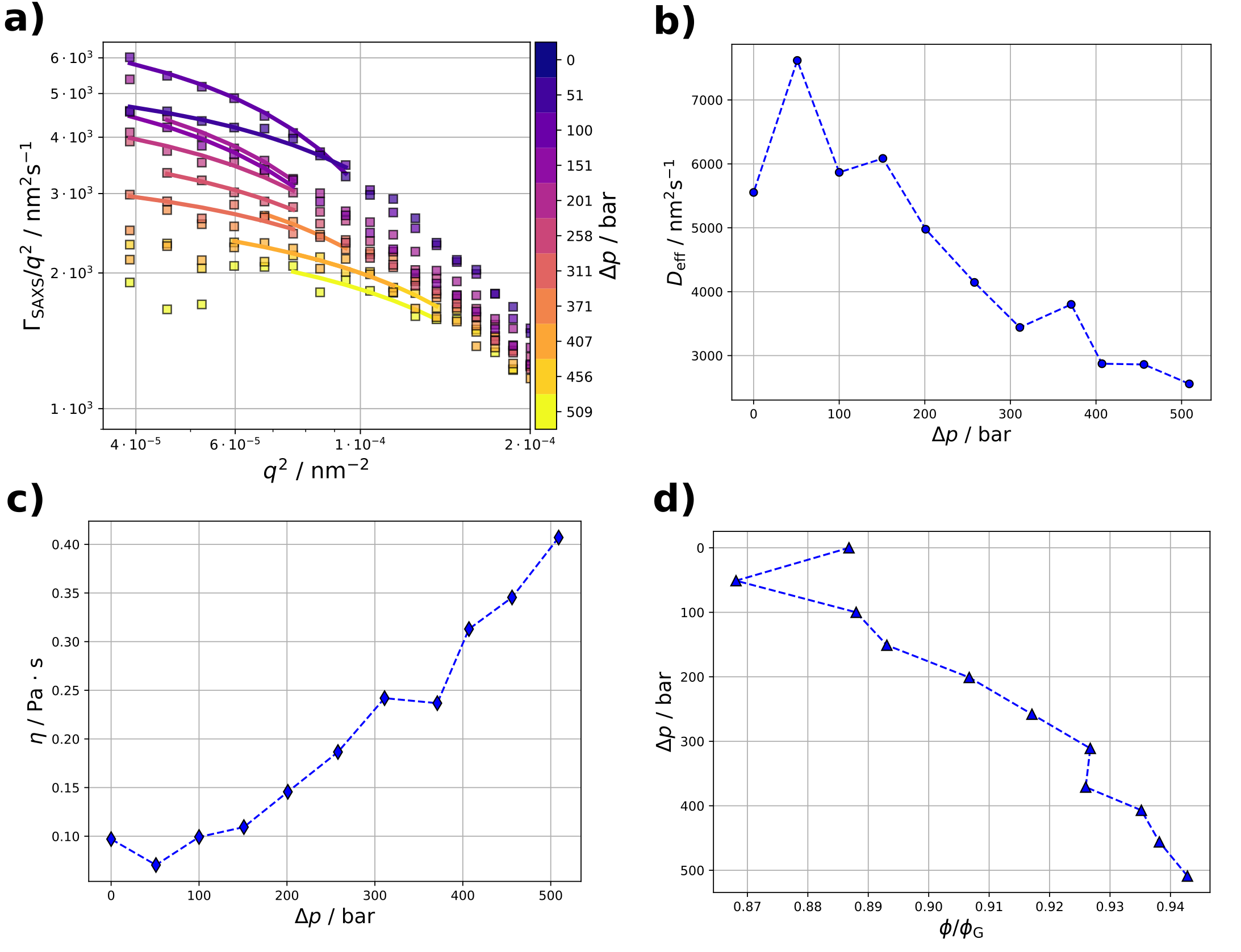}
    \caption{a) Cahn-plot for different quench depths. The solid lines represents fits of the growth rate according to the linearized Cahn-Hilliard equation. From these fits, we extracted an effective diffusion coefficient $D_{\mathrm{eff}}$ which is shown in b). Using Stokes-Einstein equation with a radius of the domain length (see Fig.\ref{fig:SAXS-ana}(c)) early after the onset of LLPS, we estimated the viscosity $\eta$. c) calculated viscosity and d) relative concentration as a function of quench depth.}
    \label{fig:saxs_gamma}
\end{figure}

\begin{figure}[!htb]
    \centering
    \includegraphics[scale=.8]{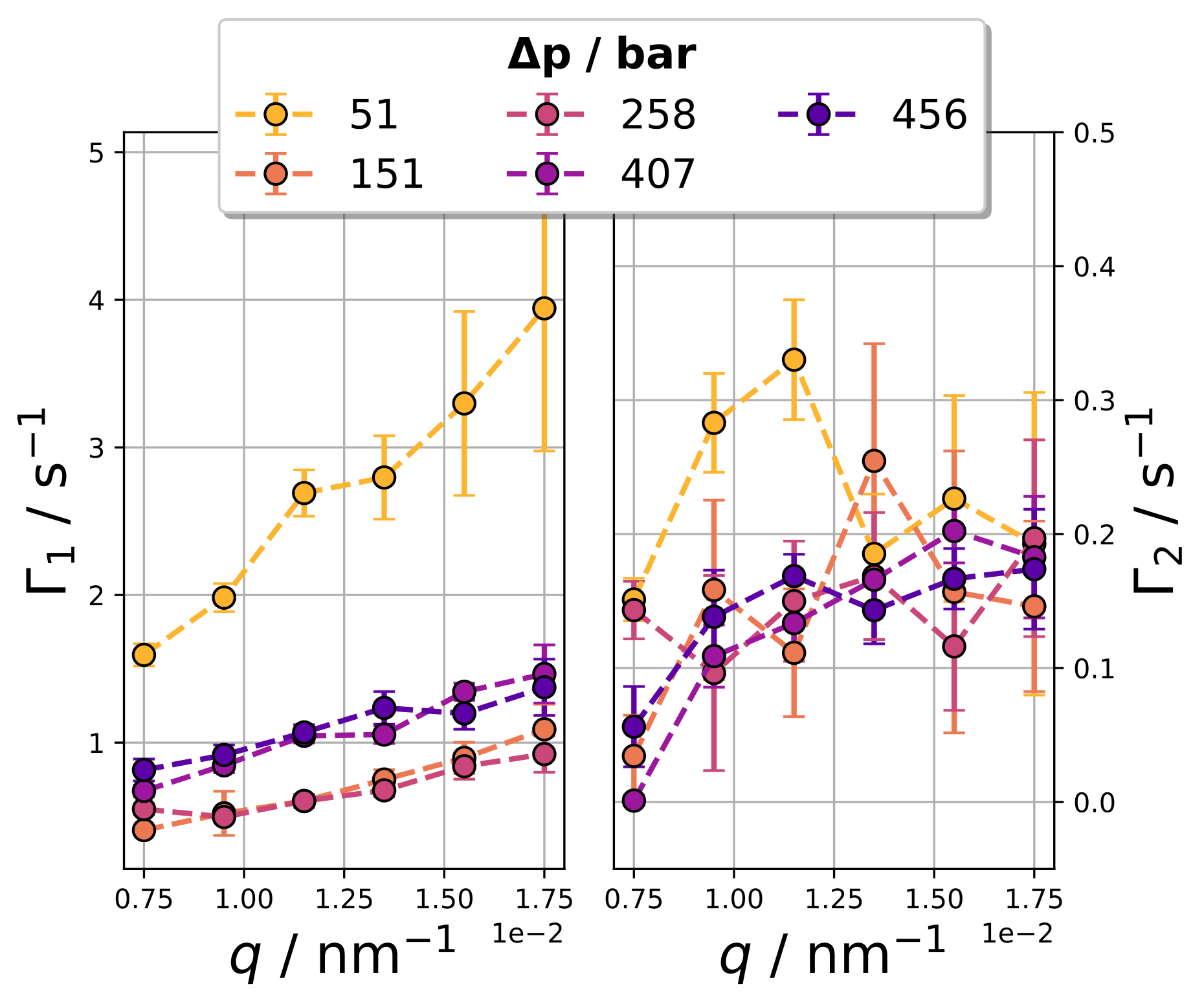}
    \caption{$q$-dependence of the decay rates $\Gamma$ for the fast (left) and slow (right) decay for elevated quench depths.}
    \label{fig:gamma_vs_q}
\end{figure}

\begin{figure}[!htb]
    \centering
    \includegraphics[scale=.8]{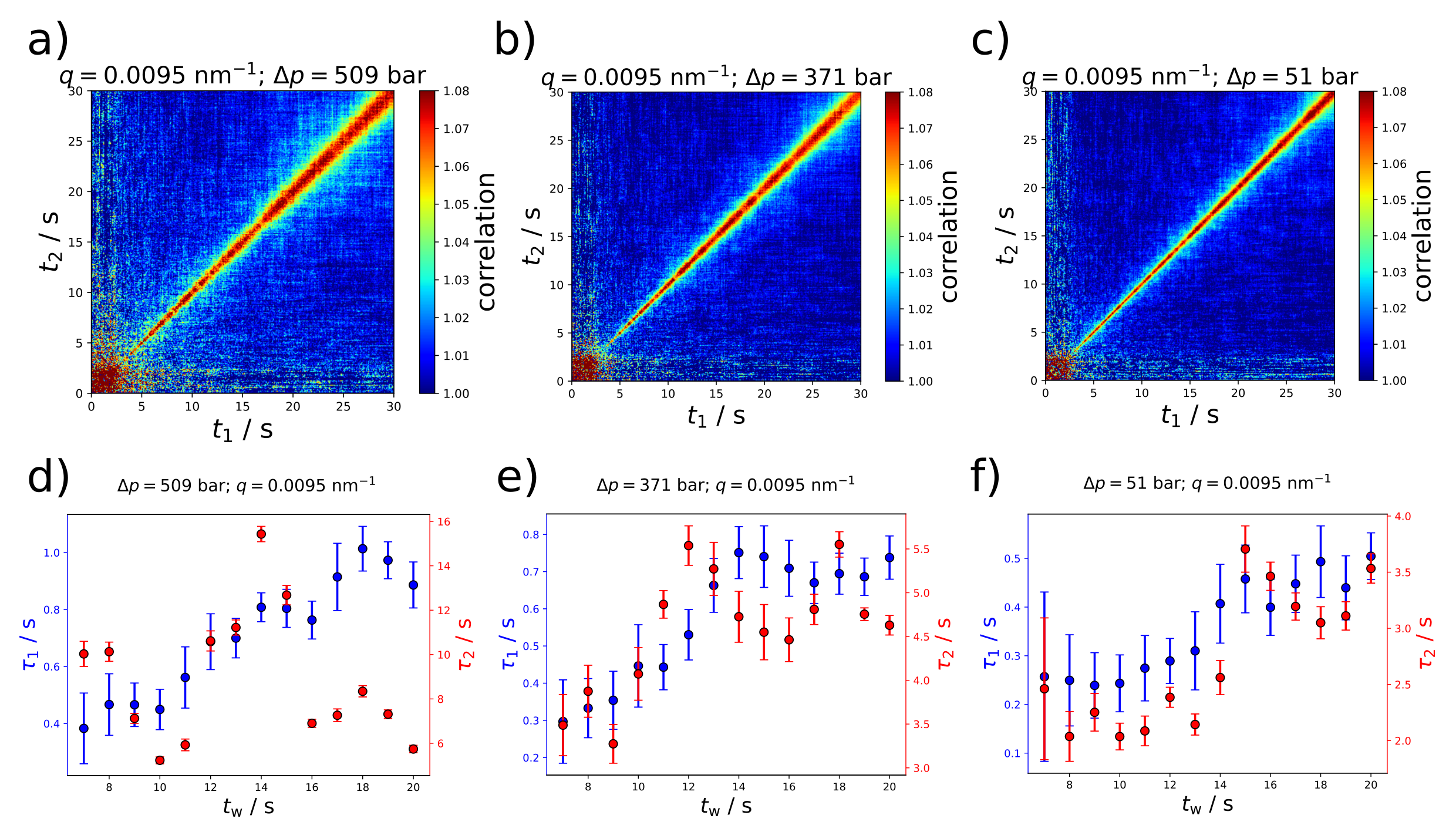}
    \caption{(a-c): two-time correlation functions for elevated quench depths at $q=\SI{0.0095}{\per\nano\meter}$. (d-f): Corresponding relaxation times for the fast (blue) and slow (red) process at the same momentum transfer.}
    \label{fig:ttc_and_tau}
\end{figure}

\end{suppinfo}



\end{document}